\pdfoutput=1
\documentclass[12pt,a4paper]{article}

\usepackage[hmargin={2cm,2cm},vmargin={2cm,2cm}]{geometry}
\usepackage[utf8]{inputenc}
\usepackage[skip=12pt,indent=12pt]{parskip}
\usepackage{amsmath}
\usepackage{amssymb}
\usepackage{authblk}
\usepackage{graphicx}
\usepackage{lineno}
\usepackage{booktabs}
\usepackage[authoryear,round]{natbib}
\usepackage{hyperref}
\usepackage{tcolorbox}
\usepackage[dvipsnames]{xcolor}

\title{Voluntary nonpharmaceutical intervention as a feedback-evolving game}

\author[1,2]{Ryuichi Kumata\thanks{ryuichi.kumata@cefe.cnrs.fr}}
\author[2,3]{Hiromu Ito}
\author[4,5,6,2,7,8]{Masato Yamamichi}
\affil[1]{CEFE, Univ Montpellier, CNRS, EPHE, IRD, France}
\affil[2]{Institute of Tropical Medicine, Nagasaki University, Japan}
\affil[3]{Multidisciplinary Pandemic Research Center, Nagasaki University, Japan}
\affil[4]{National Institute of Genetics, Japan}
\affil[5]{SOKENDAI, Japan}
\affil[6]{The University of Queensland, Australia}
\affil[7]{Yokohama National University, Japan}
\affil[8]{The University of Tokyo, Japan}
\date{}

\begin{document}

\maketitle

\begin{abstract}
    Nonpharmaceutical interventions (NPIs), such as mask wearing, can reduce epidemic spread, but their effectiveness depends on whether people adopt and sustain them. The incentive to adhere changes with infection risk and with the behavior of others. Yet it remains unclear how epidemiological protection and social reinforcement determine the strategic structure of infection control, and how epidemic feedback changes that structure over time. Here, we develop a socio-epidemiological model in which prevalence changes behavioral incentives while adherence changes transmission, separating private protection from source control and incorporating conformity. At a given prevalence, private protection favors adherence, source control creates opportunities for free-riding, and conformity reinforces whichever behavior is already common, generating distinct adherence games. Because adherence reduces transmission and prevalence, the population can move between these games as the epidemic unfolds. This feedback can stabilize partial adherence, create alternative behavioral norms, or generate recurrent epidemic-behavior cycles. Moreover, interventions with the same population-level effect on transmission at a given adherence level can produce different long-term outcomes when protection is distributed differently between the individual and others. Thus, voluntary infection control is a feedback-evolving game whose outcome depends not only on epidemiological efficacy but also on who is protected and how behavior is socially reinforced.
\end{abstract}

\begin{tcolorbox}[width=\linewidth]
\textbf{Significance statement}\\
During epidemics, people change protective behavior as infection risk and the behavior of others change. Their collective behavior, in turn, changes transmission and therefore the incentives to keep protecting themselves and others. We show that this feedback can change the game of NPI as an epidemic unfolds. Protection of the individual promotes adherence, whereas protection provided to others can create opportunities for free-riding. Social conformity reinforces whichever behavior is already common. Together, these forces can produce partial adherence, alternative social norms, and recurrent epidemic–behavior cycles. Consequently, interventions with the same effect on transmission at a given level of adherence can have different long-term outcomes depending on who receives the protection and how behavior is socially reinforced.
\end{tcolorbox}

\newpage


\section*{Introduction}

During epidemics of infectious disease such as COVID-19 and influenza, individuals may change their behavior to reduce infection risk and transmission, which is called nonpharmaceutical intervention (NPI), including by wearing masks, reducing close contacts, or avoiding high-risk social interactions \citep{funk2010modelling,traulsen2023individual}. NPI can provide two forms of protection. They can reduce infection risk for the individual who adopts them, and they can reduce transmission from that individual when infected. We call these effects private protection and source control, respectively. Many NPI behaviors combine these two forms of protection, but their relative contributions vary among behaviors. Face masks provide a clear example: experimental and empirical studies of COVID-19 and influenza show that masks can reduce exposure when worn by a susceptible individual and transmission when worn by an infectious individual, with these effects depending on mask type and fit \citep{ueki2020effectiveness,koh2022outward,leith2021quantitative,leung2020respiratory}. 

These differences in NPI efficacy affect the incentive for voluntary adherence. Private protection directly reduces the infection risk of adherers and therefore provides a direct incentive to adhere. In contrast, the source-control effect of adherence primarily benefits others by reducing onward transmission from adherers. Non-adherers can therefore benefit from source control effects provided by adherers without paying the cost of adherence. This creates an opportunity for non-adherers to free-ride on protection provided by adherers. Game-theoretic studies have revealed the possibility that the efficacy of vaccination is undermined by individual-level decision making because of incentives to free-ride \citep{bauch2004vaccination,bauch2005imitation}. More recent work has shown that the balance between protection of the user and protection of others can determine the resulting social dilemma underlying NPI behaviors \citep{pastor2022advantage,traulsen2023individual,glaubitz2024social}. Thus, the epidemiological effects of an NPI behavior influence not only how well it controls infection, but also whether individuals have an incentive to adopt and maintain it.

Adherence to NPI behaviors also depends on the social environment \citep{heiman2023descriptive}. During the COVID-19 pandemic, mask wearing was influenced by whether the behavior was perceived as common or socially expected \citep{heiman2023descriptive,yang2022sociocultural}. Similar social influences have been reported for other infection-control behaviors, including vaccination and handwashing \citep{bruine2019reports,mussio2022healthy,grover2018social}. One mechanism that can generate such frequency dependence is conformity, broadly referring to a tendency to favor behaviors that are already common \citep{efferson2008conformists}. We use the term conformity in the minimal sense of a frequency-dependent incentive to match the behavior of others. Such social reinforcement can maintain adherence when adherence is common, but can also discourage it when non-adherence is common \citep{efferson2008conformists}. Previous studies have shown that social norms can affect epidemiological dynamics by shaping the uptake of NPI \citep{oraby2014influence,qiu2022understanding,morsky2023impact,iwasa2023waves}. Taken together, the incentive to adopt an infection-control behavior depends on both epidemiological and social factors: how the behavior protects the individual and others, the current infection risk, and the behavior of others.

A key feature of human epidemic dynamics is that these behavioral incentives are not fixed. Human behavior changes epidemiological dynamics, while epidemiological dynamics in turn change the incentives for behavior \citep{bergstrom2024human}. As infections rise, protective behaviors become more valuable and adherence may increase; as adherence increases, transmission falls and the infection risk that motivated adherence declines. Models of coupling epidemiological dynamics to adaptive changes in human behaviors have revealed that this feedback can reshape epidemic trajectories and long-term disease burden \citep{funk2010modelling,arthur2021adaptive,qiu2022understanding,saad2023dynamics,glaubitz2024social}. The adherence game is therefore not fixed: as prevalence changes, the same infection-control behavior can move between different strategic situations. This links voluntary epidemic control to evolutionary games with environmental feedback, in which behavior changes the environment that in turn changes behavioral incentives \citep{weitz2016oscillating,tilman2020evolutionary,ito2024complete}. Recent work has also coupled behavioral dynamics to interventions that reduce susceptibility and/or transmission, showing that these different routes of protection can interact with epidemic feedback to shape adherence and prevalence \citep{martin2026wearing}. Yet it remains unclear how epidemiological protection and social reinforcement determine the strategic structure of infection control at a given epidemic state, and how epidemic feedback changes that structure as the epidemic unfolds.

Here, we develop a minimal socio-epidemiological model of voluntary adherence to NPI behaviors that separates private protection from source control and includes conformity. We show that these mechanisms generate distinct adherence games at a given prevalence and that epidemic feedback causes the strategic structure itself to evolve over time. The resulting dynamics include stable partial adherence, multistability, and recurrent epidemic–behavior cycles. Moreover, interventions with the same population-level effect on transmission can produce different long-term outcomes when their protective effects are distributed differently between the individual and others.

\section*{Model framework}

To formalize the socio-epidemiological feedback underlying NPI adherence, we build on the behavioral--epidemiological framework of Saad-Roy and Traulsen \citep{saad2023dynamics}, in which SIRS epidemic dynamics are coupled to payoff-biased social learning of NPI adherence. We retain its basic epidemic--behavior feedback while extending the intervention and payoff structures in two directions.

\begin{figure}
    \centering
    \includegraphics[width=1\linewidth]{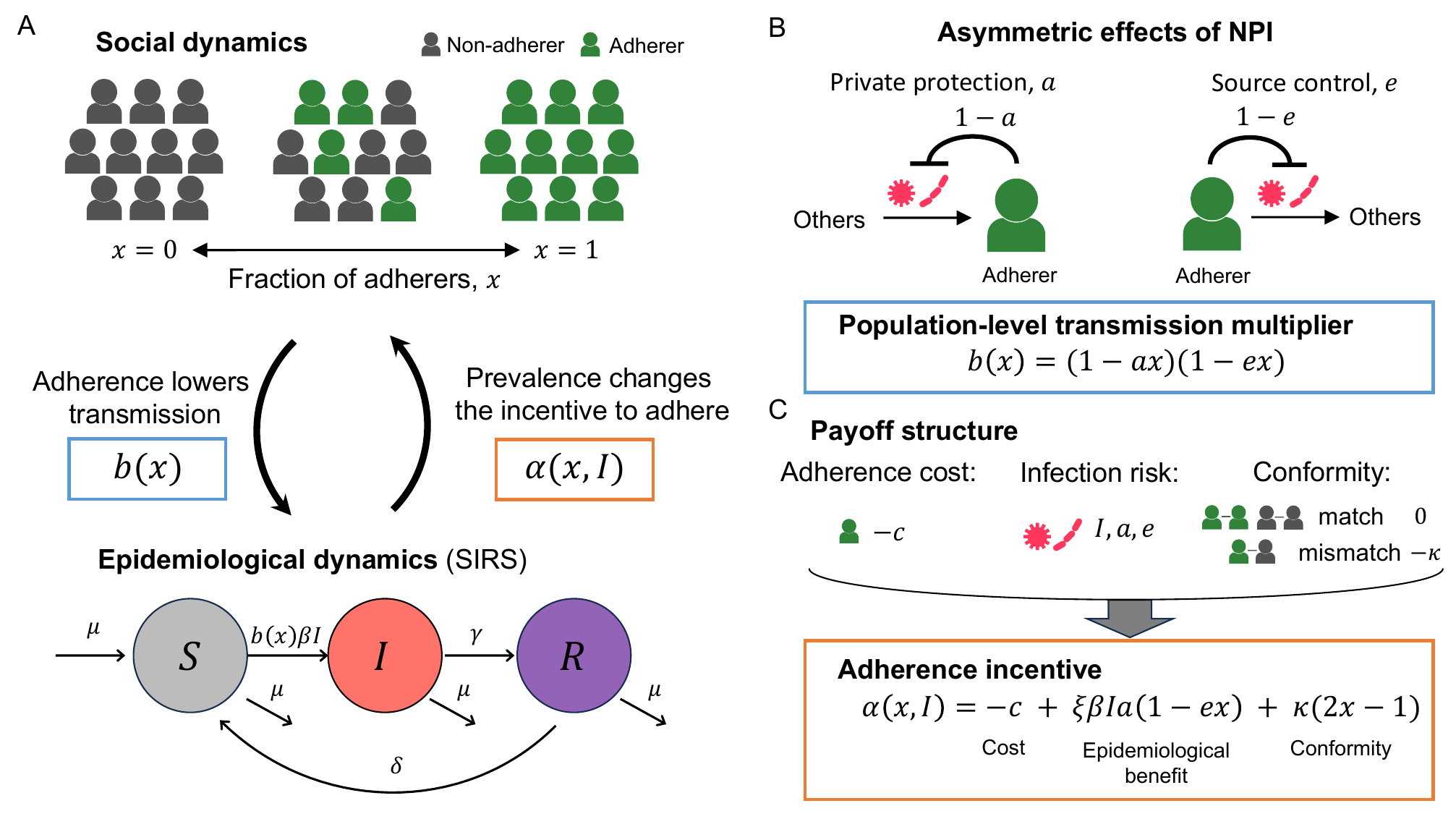}
    \caption{\textbf{Coupled social and epidemiological dynamics of voluntary NPI adherence.} (A) The fraction of adherers, $x$, changes through payoff-biased social learning and is coupled to SIRS epidemic dynamics. Adherence reduces the population-level transmission multiplier, $b(x)$, whereas infection prevalence, $I$, changes the incentive to adhere, $\alpha(x,I)$. (B) An NPI can provide private protection, $a$, by reducing infection acquisition by an adherer, and source control, $e$, by reducing onward transmission from an adherer. Under homogeneous mixing, their combined population-level effect is $b(x)=(1-ax)(1-ex)$. (C) The adherence incentive combines the cost of adherence, its private epidemiological benefit, and a conformity payoff from matching the behavior of others: $\alpha(x,I)=-c+\xi\beta I a(1-ex)+\kappa(2x-1)$.
}
    \label{fig:model}
\end{figure}

First, we distinguish two components of NPI effects: private protection and source control (Fig.~\ref{fig:model}B). Let $x\in[0,1]$ denote the fraction of individuals adhering to the NPI. Private protection, measured by $a\in[0,1]$, reduces the susceptibility of an adherer, whereas source control, measured by $e\in[0,1]$, reduces transmission from an adherer to others. Assuming homogeneous mixing and the same adherence frequency among susceptible and infected individuals, the population-level transmission multiplier, $b(x)$, is
\begin{equation}
b(x)=(1-ax)(1-ex).
\label{eq:b}
\end{equation}
Thus, private protection and source control both reduce population-level transmission, but they allocate protection differently between the actor and other individuals (see Eq.~\eqref{eq:payoff_matrix}). For minimality, we assume state-blind adherence: individuals use the same behavioral strategy regardless of their current epidemiological state, so the adherence frequency $x$ is shared across susceptible and infected individuals.

Second, we introduce social conformity into individual decision-making. Individuals choose between adherence, $A$, and non-adherence, $N$. Adherence entails a cost $c>0$. At infection prevalence $I$ (fraction of infected individuals), infection generates a perceived loss $\xi\beta I$, where $\xi$ measures the perceived harm of infection and $\beta$ is the baseline transmission rate. Individuals additionally incur a social mismatch cost due to a social norm ($\kappa\geq0$) when their behavior differs from that of their interaction partner.

The above assumption yields the following payoff matrix of the focal row player:
\begin{equation}
\Pi(I) =
\bordermatrix{
      & N & A \cr
N     & -\xi\beta I
      & -\xi\beta I(1-e) -\kappa \cr
A     & -c-\xi\beta I(1-a)-\kappa
      & -c-\xi\beta I(1-a)(1-e) \cr
}.
\label{eq:payoff_matrix}
\end{equation}
When two non-adherers interact, the focal individual experiences the full infection loss, $-\xi\beta I$. If the focal individual adheres while its partner does not, private protection reduces its infection risk by $1-a$, but it pays both the adherence cost $-c$ and the mismatch cost $-\kappa$. Conversely, if only the partner adheres, the focal individual benefits from source control with the reduction of infection risk by $1-e$ but incurs the mismatch cost $-\kappa$.

Averaging over encounters with adherers and non-adherers gives the expected payoff of each strategy. The difference between these payoffs then gives the net incentive to adhere,
\begin{align}
\alpha(x,I)
&=
-c+a\xi\beta I(1-ex)+\kappa(2x-1).
\label{eq:adherence_incentive}
\end{align}
The three terms in Eq.~\eqref{eq:adherence_incentive} represent the cost of adherence, its private epidemiological benefit, and the frequency-dependent incentive generated by social matching, respectively (Fig.~\ref{fig:model}C). Higher prevalence and stronger private protection increase the incentive to adhere. In contrast, source control provided by other adherers reduces the marginal private benefit of adherence by the factor $(1-ex)$. Social matching disfavors adherence when it is rare ($x<1/2$) and favors it when it is common ($x>1/2$).

We model payoff-based social learning using the replicator equation \citep{hofbauer1998evolutionary}. Let $S$, $I$, and $R$ denote the susceptible, infected, and recovered fractions, respectively, with $S+I+R=1$. Taking $R=1-S-I$, the coupled socio-epidemiological dynamics are
\begin{align}
\frac{dS}{dt}
&=
\mu-\beta b(x)SI+\delta(1-S-I)-\mu S,
\label{eq:coupled_S}\\
\frac{dI}{dt}
&=
\beta b(x)SI-(\gamma+\mu)I,
\label{eq:coupled_I}\\
\frac{dx}{dt}
&=
\epsilon x(1-x)\alpha(x,I),
\label{eq:coupled_x}
\end{align}
where $\gamma$ is the recovery rate, $\mu$ is the demographic turnover rate, $\delta$ is the rate of waning immunity, and $\epsilon$ controls the rate of behavioral updating relative to epidemiological dynamics.

Equations~\eqref{eq:coupled_S}--\eqref{eq:coupled_x} define the socio-epidemiological feedback shown in Fig.~\ref{fig:model}A. Adherence reduces transmission through $b(x)$, whereas epidemic prevalence changes the incentive to adhere through $\alpha(x,I)$. Derivations of the expected payoffs and the complete classification of the instantaneous game are provided in SI Appendix, Sections S1 and S2.

\section*{Results}

\subsection*{Epidemiological protection and conformity determine the instantaneous adherence game}

\begin{figure}
    \centering
    \includegraphics[width=0.8\linewidth]{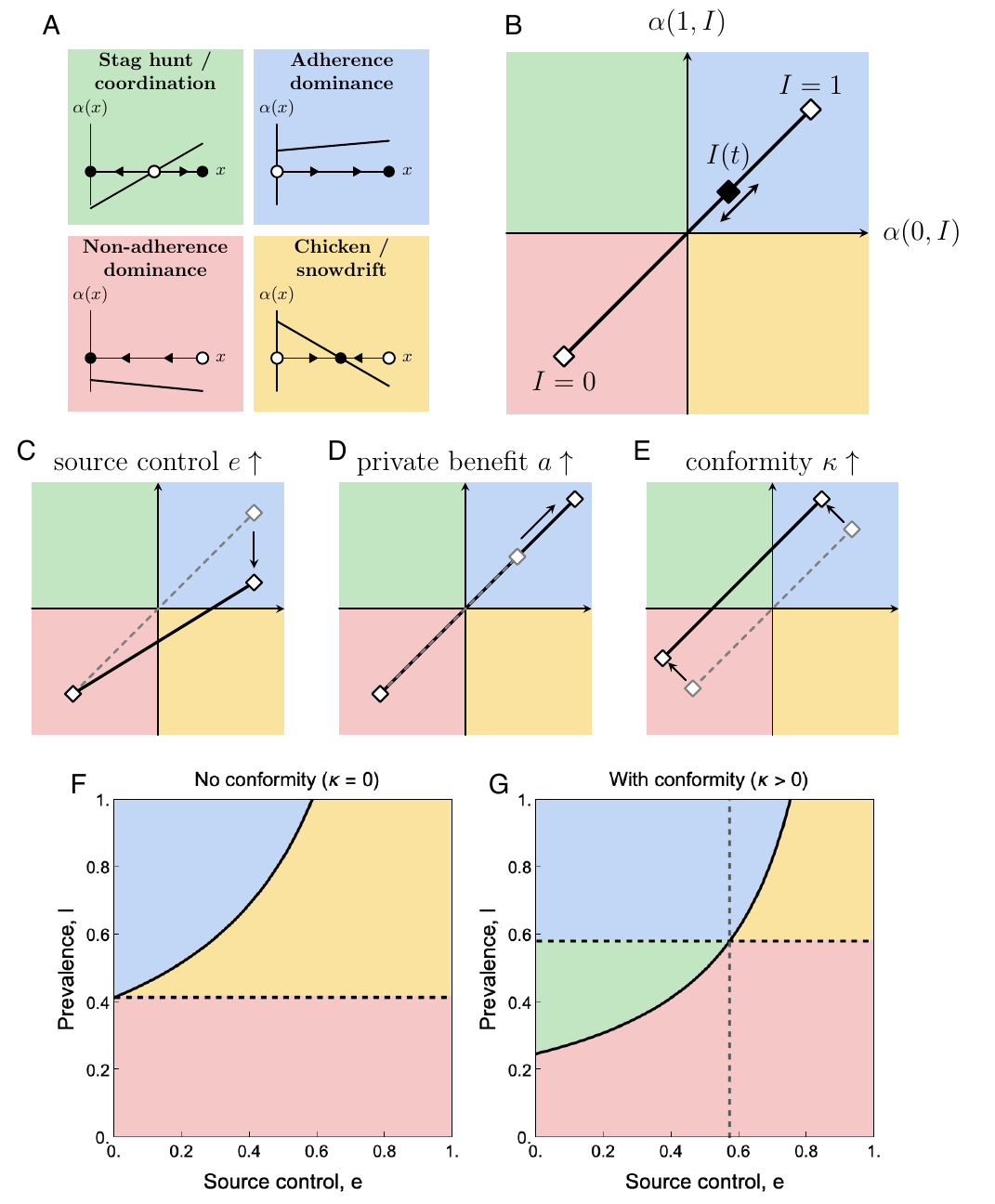}
    \caption{\textbf{Game structure of voluntary infection control} (A) Four strategic structures classified by the adherence incentives when adherence is rare, $\alpha(0,I)$, and common, $\alpha(1,I)$. Curves show $\alpha(x,I)$, arrows indicate behavioral change, and filled and open circles denote stable and unstable equilibria. (B) Game-space representation of the instantaneous adherence game without source control or conformity ($e=0$ and $\kappa=0$). Varying prevalence $I$ traces a straight trajectory in the $(\alpha(0,I),\alpha(1,I))$ plane; open diamonds mark $I=0$ and $I=1$, and the filled diamond marks the current game at $I(t)$. Colors indicate the four game types in (A). (C–E) Increasing source control $e$ rotates the trajectory toward the chicken/snowdrift region, increasing private protection $a$ extends it, and increasing conformity $\kappa$ shifts it toward the coordination region. Black trajectories and diamonds indicate the game after each parameter increase. (F,G) Instantaneous game structure across source control $e$ and prevalence $I$ without (F) and with (G) conformity. Dashed and solid boundaries satisfy $\alpha(0,I)=0$ and $\alpha(1,I)=0$, respectively; the vertical dashed line in (G) marks $e=2\kappa/(c+\kappa)$. Panels F–G use $a\xi\beta/c=2.4$, with $\kappa/c=0$ in F and $0.4$ in G.
}
    \label{fig:games}
\end{figure}

We first assumed that infection prevalence, $I$, is fixed and examined how epidemiological protection and conformity structure the instantaneous game of NPI adherence. Because the adherence incentive is linear in the adherence frequency $x$, the game is completely classified by the incentive to adhere when adherence is rare,
\begin{equation}
\alpha(0,I)
=
-c+a\beta\xi I-\kappa,
\label{eq:alpha0}
\end{equation}
and when adherence is common,
\begin{equation}
\alpha(1,I)
=
-c+a(1-e)\beta\xi I+\kappa.
\label{eq:alpha1}
\end{equation}
These quantities determine whether rare adherence can invade and whether widespread adherence can resist invasion by non-adherers. Their signs generate four canonical two-strategy games, corresponding to dominance, anti-coordination (chicken/snowdrift), and coordination (stag-hunt) (Fig.~\ref{fig:games}A) \citep{doebeli2005models,ito2024complete}. If both are negative, non-adherence is dominant; if both are positive, adherence is dominant. If $\alpha(0,I)>0$ but $\alpha(1,I)<0$, adherence spreads when rare but is undermined by free-riding when common, producing a chicken- or snowdrift-like game. Conversely, if $\alpha(0,I)<0$ but $\alpha(1,I)>0$, adherence cannot spread when rare but can be maintained once established, producing a stag-hunt or coordination-like game.

The classification based on two endpoint incentives also provides a geometric representation of the instantaneous game. At any prevalence $I$, the game corresponds to a point ($\alpha(0,I), \alpha(1,I)$) in game space (Fig.~\ref{fig:games}B). As prevalence varies, this point forms the line
\begin{align}
    G(I) = G(0) + I v,
\end{align}
where $G(0)=(-c-\kappa,-c+\kappa)$ and $v=a\xi\beta(1,1-e)$. $G(1)$ is the game structure at the theoretical maximal prevalence, $I=1$.
Thus, depending on the infection prevalence, the population can move between different game structures. Figure~\ref{fig:games}B depicts the game structures when both conformity and source control are absent ($e=\kappa=0$) as a baseline setting. In the baseline scenario shown in Figure~\ref{fig:games}B, the trajectory lies on the diagonal and passes through the origin, meaning that increasing prevalence shifts the game directly from non-adherence dominance to adherence dominance .

The game geometry makes clear the distinct roles of source control, private protection, and conformity (Fig.~\ref{fig:games}C–E). Source control $e$ changes the direction of the trajectory. It leaves $\alpha(0,I)$ unchanged but lowers $\alpha(1,I)$, because a non-adherer can benefit from source control supplied by adherers when adherence is common. Increasing $e$ therefore rotates the trajectory clockwise, toward the chicken/snowdrift region (Fig.~\ref{fig:games}C). Private protection $a$ controls how strongly the game responds to prevalence. When $I=0$, protection has no epidemiological value, so changing $a$ leaves the $I=0$ endpoint unchanged. As prevalence rises, however, stronger private protection increases the benefit of avoiding infection and therefore moves the game farther from this endpoint. Geometrically, increasing $a$ stretches the game trajectory without changing its direction (Fig.~\ref{fig:games}D). Conformity $\kappa$ acts differently. It lowers the incentive to adhere when adherence is rare but raises it when adherence is common, shifting the entire trajectory toward the upper-left, coordination region without changing its direction (Fig.~\ref{fig:games}E). Thus, while private protection amplifies the prevalence-dependent incentive to adhere without changing the direction of the game trajectory, source control promotes stable partial adherence through negative frequency dependence, and conformity generates positive frequency dependence. 

The endpoint incentives define the prevalence thresholds for invasion and maintenance of NPI behavior (Fig.~\ref{fig:games}F,G). If infection prevalence is greater than the invasion threshold $I_{0}=(c+\kappa)/(a\beta\xi)$ (horizontal dashed lines in Fig.~\ref{fig:games}F,G), rare adherers invade a population in which most individuals do not adhere. If infection prevalence is greater than the maintenance threshold $I_{1}=(c-\kappa)/(a(1-e)\beta\xi)$ (curves in Fig.~\ref{fig:games}F,G), then the rare non-adherers cannot invade when adherence is common.

The ordering of these thresholds determines how conformity and source control shape the game structure. Without conformity, invasion precedes maintenance for any $e>0$ (Fig.~\ref{fig:games}F). Increasing prevalence therefore shifts the game from non-adherence dominance through a chicken/snowdrift game to adherence dominance, with stronger source control widening the intermediate free-riding region. Conformity can reverse the ordering of these transitions (Fig.~\ref{fig:games}G). The two boundaries coincide at a critical source control value $e_c=2\kappa/(c+\kappa)$. For $e<e_c$, the intermediate game is coordination-like, whereas for $e>e_c$, it is chicken/snowdrift-like.

When the mismatch cost exceeds the cost of adherence, $\kappa>c$, widespread adherence is socially stable even when its private epidemiological benefit is small (Fig.~S2). The game is then coordination-like at low private benefit and becomes adherence-dominant as that benefit increases. Strong conformity therefore does not favor adherence itself; it strengthens the positive frequency dependence of adherence and stabilizes whichever behavior is already common.

\subsection*{Epidemic feedback transforms adherence games into distinct dynamics}

The preceding analysis investigated the game structure underlying voluntary infection control given a fixed infection prevalence. In the coupled system, however, prevalence is itself changed by adherence. Increasing adherence reduces transmission and moves the instantaneous game toward lower $I$, whereas declining adherence raises prevalence again and moves the game in the opposite direction. The strategic environment of the infection-control behavior therefore evolves together with the epidemic states.

This feedback introduces a second route through which collective adherence can weaken its own incentive. Source control acts directly at a given prevalence: as adherence increases, non-adherers receive more protection from adherent partners, reducing the marginal benefit of adhering. Epidemic feedback acts indirectly: greater adherence lowers prevalence, which in turn reduces the private value of protection. This prevalence-mediated self-limitation is closely related to the mechanism underlying free-riding in vaccination games \citep{bauch2005imitation}.

To visualize this feedback, we plotted the endemic prevalence at a fixed adherence frequency, $I=E(x)$, together with the interior behavioral nullcline $I=B(x)$.
For fixed $I$, the stability of an interior behavioral equilibrium is determined by the frequency dependence of the adherence incentive. Because $\partial \alpha/\partial I>0$, this stability is equivalently indicated by the slope of $B(x)$: $B'(x)>0$ corresponds to $e>e_c$ and a chicken/snowdrift-like intermediate game, whereas $B'(x)<0$ corresponds to $e<e_c$ and a coordination-like intermediate game.

Interior stationary states occur at intersections of $I=E(x)$ and $I=B(x)$. Boundary stationary states occur at $x=0$ or $x=1$, together with the corresponding epidemiological equilibrium. Local stability of these states is determined from the Jacobian of the full $(S,I,x)$ system.

\begin{figure}
    \centering
    \includegraphics[width=1\linewidth]{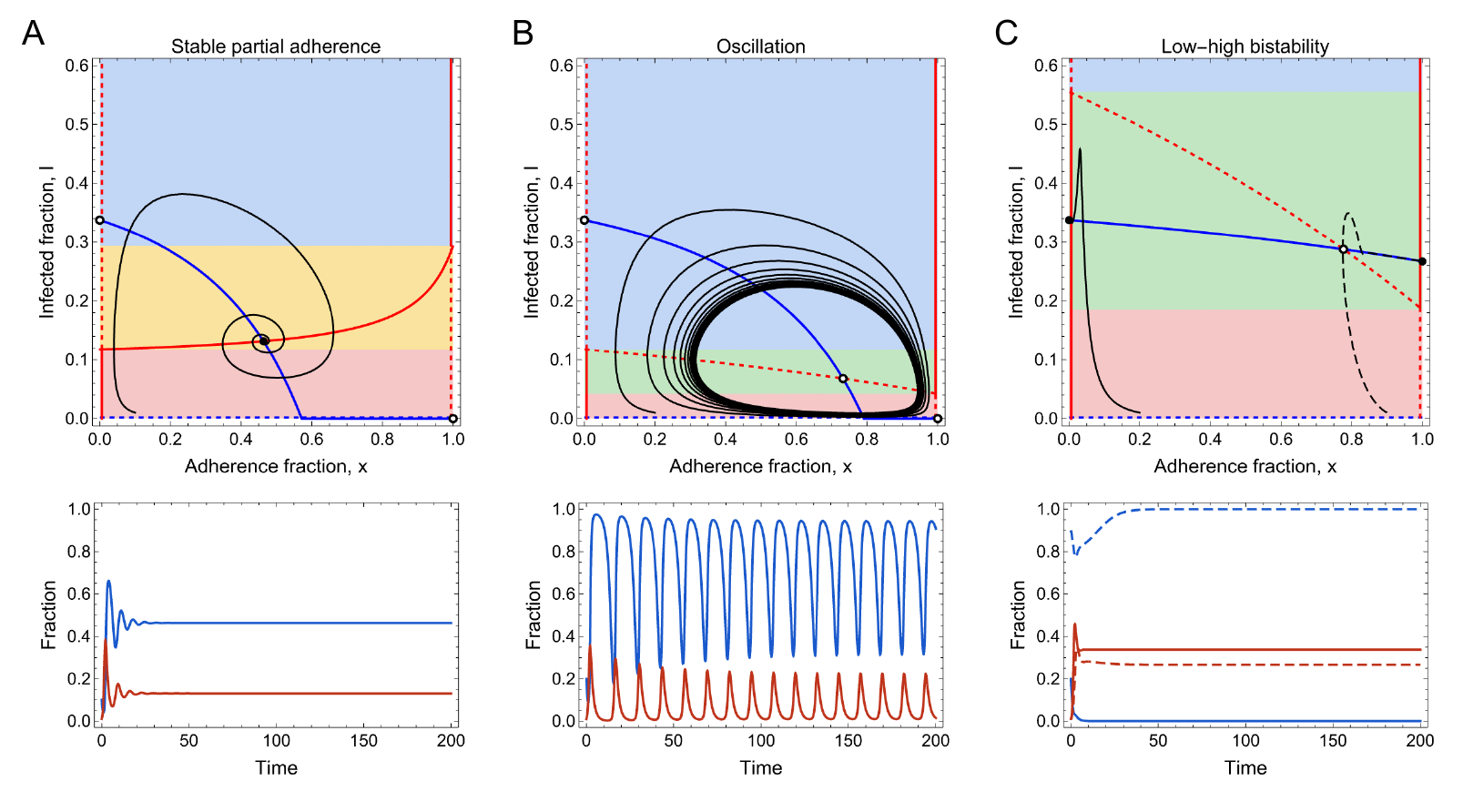}
    \caption{ \textbf{Representative dynamics generated by socio-epidemiological feedback.} Upper panels show trajectories of the full $(S,I,x)$ system projected onto the adherence--prevalence plane. Blue and red curves denote the endemic equilibrium $I=E(x)$ and the behavioral nullcline $I=B(x)$, respectively; background colors indicate the instantaneous game structure. Filled and open circles denote stable and unstable stationary states. Lower panels show the corresponding time series of adherence, $x$ (blue), and prevalence, $I$ (red). (A) Stable partial adherence ($a=0.85,e=0.9$). (B) A stable epidemic--behavior cycle around an unstable interior equilibrium  ($a=0.85,e=0.3$). (C) Bistability between no adherence and full adherence  ($a=0.18,e=0.25$); solid and dashed trajectories begin from different initial conditions. The dashed trajectory overlaps the $x$-nullcline near the convergence to full adherence ($x=1$). Fixed parameters are $\beta=4.0$, $\gamma=0.98$, $\mu=0.02$, $\delta=0.782$, $c=1.0$, $\xi=4$, $\epsilon=1.0$, and $\kappa=0.6$.}
    \label{fig:dynamics}
\end{figure}

\paragraph{Stable partial adherence.}
When epidemic-mediated negative feedback dominates social reinforcement, the system converges to a stable interior state (Fig.~\ref{fig:dynamics}A). A decline in adherence increases prevalence, which raises the private benefit of protection and restores adherence. Conversely, increasing adherence suppresses prevalence and weakens the incentive to remain adherent. This self-limiting feedback maintains both endemic infection and partial adherence.

Private protection and source control contribute to this feedback in related but distinct ways. Both reduce prevalence and thereby weaken risk-dependent adherence. Source control also reduces the private value of one's own adherence at a given prevalence because protection is supplied by other adherers.

\paragraph{Recurrent epidemic--behavior cycles.}
Restoring feedback does not always produce convergence. In Fig.~\ref{fig:dynamics}B, the interior equilibrium is unstable and the trajectory approaches a stable cycle. Rising prevalence promotes adherence, increasing adherence suppresses transmission, lower prevalence weakens adherence, and declining adherence allows infection to rebound.

The trajectory crosses several instantaneous game regions during each cycle. The oscillation therefore reflects a dynamically changing strategic environment rather than a single fixed game. It arises from the interaction between feedback strength and the finite rates of epidemiological and behavioral adjustment. The parameter dependence of this instability is examined in the next section.

\paragraph{Alternative adherence norms.}
When social reinforcement is sufficiently strong, the interior stationary state becomes a instability  threshold separating alternative boundary outcomes (Fig.~\ref{fig:dynamics}C). Initial conditions with lower adherence converge to no adherence (solid lines), whereas the higher initial adherence leads to full adherence (dashed lines). The same epidemiological parameters can therefore generate contrasting long-term social norms, with the realized outcome determined by the initial epidemiological and behavioral state. This bistability generates history-dependent outcomes and is the dynamic expression of the underlying coordination structure.

\subsection*{Conformity reorganizes the landscape of stable states}

\begin{figure}
    \centering
    \includegraphics[width=1\linewidth]{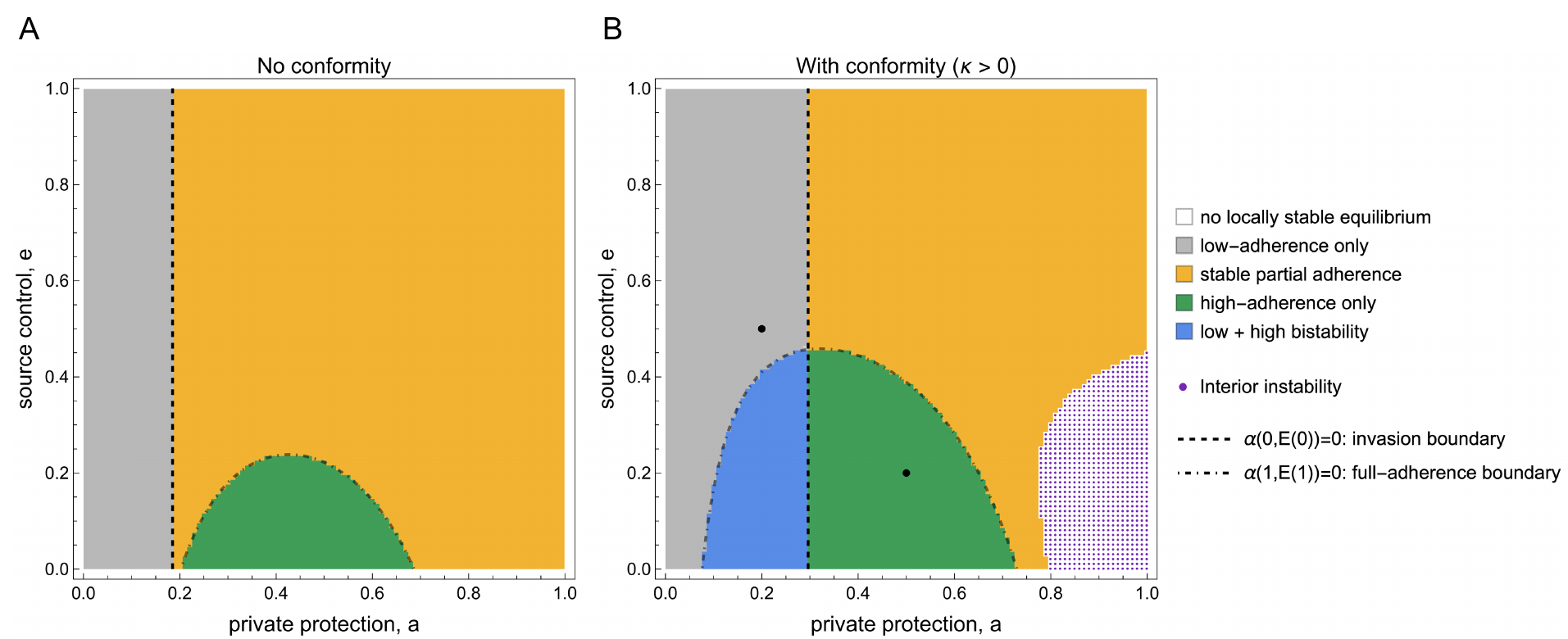}
    \caption{\textbf{Conformity reorganizes the landscape of stable socio-epidemiological states.} Locally stable stationary states across the private-protection--source-control plane under (A) no conformity, $\kappa=0$, (B) conformity, $\kappa=0.6$. Colors and patterns indicate the stable state or combination of stable states, as defined in the legend; gray denotes parameter combinations with no stable stationary state detected. Dashed and dotted black curves mark changes in the stability of the no-adherence and full-adherence states, defined by $\alpha(0,E(0))=0$ and $\alpha(1,E(1))=0$, respectively. Purple points mark loss of stability of the interior equilibrium through a complex-conjugate eigenvalue pair. Paired markers in (B) compare $(a,e)$ and $(e,a)$, which have the same transmission multiplier at fixed adherence but occupy different stability regions.  Fixed parameters are $\beta=4.0$, $\gamma=0.98$, $\mu=0.02$, $\delta=0.782$, $c=1.0$, $\xi=4$, and $\epsilon=1.0$.}
    \label{fig:atlas}
\end{figure}

We classified the locally stable stationary states across the different combinations of protection with and without conformity (Fig.~\ref{fig:atlas}). The sign of $\alpha(0,E(0))$ and $\alpha(1,E(1))$ determines the local stability of the no-adherence and full-adherence states, respectively. At each parameter combination, we classified whether no adherence, full adherence, or an interior partial-adherence state was locally stable, including regions in which several states coexisted. We mark with purple points parameter combinations at which the interior equilibrium loses stability.

\paragraph{No conformity.}
Without conformity ($\kappa=0$), weak private protection cannot compensate for the cost of adherence, and no adherence ($x=0$) is the only stable behavioral state (Fig.~\ref{fig:atlas}A). As private protection increases, adherence becomes individually worthwhile. For relatively weak source control, full adherence ($x=1$) is stable over an intermediate range of private protection. At still higher private protection, however, adherence suppresses prevalence so strongly that its private benefit declines, and the system instead converges to partial adherence.

Increasing source control contracts the full-adherence region. Although source control improves population-level protection, it also allows non-adherers to benefit from protection supplied by adherers. In the absence of social reinforcement, partial adherence therefore occupies much of the intermediate parameter region.

\paragraph{Conformity generates bistability and persistent oscillation}
Conformity produces additional dynamical regimes in the parameter space (Fig.~\ref{fig:atlas}B). A region of bistability between no adherence and full adherence (blue area) appears at relatively weak private protection. Rare adherence cannot invade in this region, but once full adherence is established, the social cost of deviating from the majority maintains it.

When private protection is large and source control is low, the interior equilibrium loses stability, giving rise to the recurrent dynamics illustrated in Fig.~\ref{fig:dynamics}B. A bifurcation analysis with $\kappa$ as the bifurcation parameter shows that the equilibrium is stable in the absence of conformity and becomes unstable only above a positive threshold through a Hopf bifurcation (Fig.~S1C). More generally, an interior equilibrium can become unstable only when $\partial\alpha/\partial x>0$. Because $\partial\alpha/\partial x=-ae\xi\beta I+2\kappa$, this condition cannot be satisfied in the absence of conformity. The location of the oscillatory instability depends on the speed of behavioral update $\epsilon$ (Fig. S1C). Thus, behavioral time scales determine when the instability occurs, whereas positive conformity is required for recurrent dynamics in the parameter regimes examined here.

\paragraph{Strong conformity.}
When the mismatch cost exceeds the cost of adherence, $\kappa>c$, the region in which full adherence is locally stable expands substantially (Fig.~S3). Once adherence is common, the social cost of deviating can maintain it even when the private epidemiological benefit is modest. The same conformity also stabilizes non-adherence when it is already common. At weak private protection, no adherence and full adherence therefore coexist across a broad parameter region. Strong conformity thus increases history dependence rather than uniformly promoting adherence. At high source control, stable partial adherence can coexist with full adherence. Protection supplied to others preserves an opportunity for free-riding even when conformity strongly stabilizes the high-adherence state. Conformity therefore changes not only the level of adherence but also the number and type of stable outcomes.


\paragraph{Symmetric transmission effects, asymmetric behavioral outcomes.}
Comparing the two parameter combinations marked by black points in Fig.~\ref{fig:atlas}B further illustrates the asymmetry between private protection and source control. Although the two points are reflections across $a=e$ and therefore have the same transmission multiplier in Eq.~\eqref{eq:b} at any fixed adherence frequency, the private-protection-dominated intervention ($a=0.5, e=0.2$) supports full adherence, whereas the source-control-dominated intervention ($a=0.2, e=0.5$) converges to no adherence. Interventions with the same population-level transmission effect at a given adherence level can therefore produce different long-term outcomes because the allocation of effectiveness changes individual incentives.

\section*{Discussion}

Our results show that epidemiological and social mechanisms shape the game of voluntary NPI behavior, and that epidemic feedback converts these games into distinct long-term dynamics. Private protection strengthens prevalence-dependent incentives, source control shifts the game toward free-riding, and conformity shifts it toward coordination. Coupling these forces to epidemic dynamics produces stable partial adherence, multistability in adherence, and recurrent epidemic--behavior cycles. Notably, conformity can stabilize established norms but can also destabilize an interior equilibrium. Long-term outcomes therefore depend on intervention effects, social incentives, and their feedback with infection dynamics.

NPI behaviors often differ in the relative strength of private protection and source control. For masks and respirators, inward and outward protection depend on mask type, fit, and use \citep{leung2020respiratory,koh2022outward}. Other interventions, including distancing, ventilation, and isolation, also reduce transmission through different routes \citep{talic2021effectiveness,kucharski2020effectiveness}. Previous epidemic games have shown that protection of oneself and protection of others can create different conflicts between individual and collective interests \citep{bauch2004vaccination,pastor2022advantage,traulsen2023individual,glaubitz2024social}. Our results extend this point to a coupled behavioral--epidemiological system. Exchanging private protection and source control leaves the transmission multiplier unchanged at any fixed adherence frequency, yet it can produce different adherence games and different long-term outcomes. Once behavior evolves, the two cases are no longer epidemiologically equivalent because their different incentives lead to different adherence and epidemic trajectories. Thus, how an NPI behavior reduces transmission can shape the social dilemma surrounding its use and, through behavioral feedback, its long-term epidemiological outcome.

Our analysis also separates two routes through which collective adherence can weaken its own incentive and make its maintenance difficult. Source control creates a direct opportunity for free-riding: at the same prevalence, non-adherers receive more protection as more interaction partners adhere. It therefore creates negative frequency dependence and selectively weakens the maintenance of common adherence. Collective adherence also lowers transmission and prevalence, which reduces the private value of protection. This prevalence-mediated self-limitation is closely related to the free-riding mechanism emphasized in vaccination games \citep{bauch2004vaccination,bauch2005imitation,traulsen2023individual}. The two mechanisms act at different levels: source control changes payoffs at a given epidemic state, whereas epidemic feedback changes the state at which those payoffs are evaluated. 

Conformity is a common source of frequency-dependent social influence and has long been studied as an important force shaping social dynamics and cultural evolution \citep{javarone2014social,henrich1998evolution,efferson2008conformists}. Social norms are also widely implicated in behavioral responses to infectious disease, and perceived infection threat may itself increase conformist tendencies \citep{murray2012threat,bavel2020using}. Preventive behaviors such as mask wearing and influenza vaccination are associated with perceptions of others’ behavior and expectations \citep{heiman2023descriptive,bruine2019reports,mussio2022healthy}. Similar normative influences have been identified for routine hygiene behaviors such as handwashing \citep{curtis2009planned}. We represent this broad social influence in a minimal form as conformity, an incentive to match prevailing behavior. Conformity creates positive frequency dependence because either adherence or non-adherence becomes more attractive when it is common. It can therefore counter source-control-mediated free-riding and shift the adherence game toward coordination.

A broader implication is that NPI adherence need not correspond to one fixed game. Infection prevalence changes the private value of protection, and adherence changes prevalence in return. The population can therefore move among non-adherence dominance, adherence dominance, free-riding (chicken/snowdrift), and coordination as an epidemic unfolds. This provides a concrete epidemiological example of evolutionary games with environmental feedback \citep{weitz2016oscillating,tilman2020evolutionary,ito2024complete}. Here, the environment does not merely change the strength of selection; it can change which game is being played. This view also helps organize the long-term dynamics. Epidemic-mediated negative feedback can stabilize partial adherence: more adherence lowers prevalence, which then weakens adherence.
Previous models of NPI adherence have shown that partial adherence can be stable \citep{martin2026wearing,saad2023dynamics}. Our results show that social frequency dependence can qualitatively change these dynamics. Conformity can destabilize this interior equilibrium by adding positive frequency dependence, producing recurrent epidemic--behavior cycles. This result is consistent with a study of epidemiological dynamics with human behavior \citep{iwasa2023waves}. Furthermore, the game perspective clarifies why conformity is required for oscillations. Destabilization of the interior equilibrium requires positive frequency dependence, corresponding to a coordination (stag-hunt) game. Source control generates negative frequency dependence, whereas conformity promotes positive frequency dependence; sufficiently strong conformity can therefore destabilize the interior equilibrium and generate oscillations.

The model deliberately isolates these mechanisms using a minimal evolutionary game framework. We classified the strategic structure of adherence from individual incentives but did not quantify the associated loss of collective welfare. A natural extension would be to define a social optimum and quantify the social efficiency deficit \citep{arefin2020social}. Real behavioral incentives are likely to be more complex than the assumption in our model. Individuals may value the protection they provide to others, which can partly offset the free-riding incentive created by source control \citep{jordan2021don,pfattheicher2020emotional}. We represented conformity as a symmetric, population-wide matching payoff. Real social influence may instead depend on descriptive and injunctive norms, nonlinear thresholds, reference groups, and contact networks \citep{oraby2014influence,morsky2023impact,qiu2022understanding,rabb2022influence}. It may itself change with perceived infection threat \citep{murray2012threat}. We also assumed homogeneous mixing, homogeneous costs and responses, and the same adherence frequency among susceptible and infected individuals. The last assumption may be especially important for source control, because symptoms, testing, or knowledge of infection can change the incentive to protect others. These extensions may shift the boundaries between the dynamical regimes identified here. The basic point, however, is general: an NPI behavior that strongly suppresses transmission need not be easy to sustain voluntarily. Its success depends on how it protects the user and others, how those effects interact with social incentives, and how the resulting behavior changes the epidemic conditions that motivate it.

\section*{Materials and Methods}

\subsection*{Stationary states and local stability}

To analyze and classify the long-term behavior of socio-epidemiological feedback dynamics in Eqs.~\eqref{eq:coupled_S}-\eqref{eq:coupled_x}, we investigated equilibrium points of the system and determined their local stability from the eigenvalues of the Jacobian of the full $(S,I,x)$ system. Analytical expressions for the Jacobian and the local stability conditions are provided in SI Appendix, Sections S3 and S4.

As the adherence $x$ depends only on infection prevalence $I$, the projection of the dynamics to $x$-$I$ plane gives convergence to the long-term behavior and provides the geometrical interpretation of the feedback (Fig.~\ref{fig:dynamics}).
For a fixed adherence frequency $x$, the endemic infection prevalence (blue line and curve in upper panels Fig.~\ref{fig:dynamics}) is
\begin{align}
E(x)
&=
\begin{cases}
K\left(1-\dfrac{1}{\rho b(x)}\right),
& \rho b(x) > 1,\\[6pt]
0,
& \rho b(x) \le 1,
\end{cases}
\label{eq:endemic_branch}
\end{align}
where $\rho=\beta/(\gamma+\mu)$ is the basic reproductive number and $b(x)=(1-ax)(1-ex)$. The interior behavioral nullcline (red curve in Fig.~\ref{fig:dynamics}) is given by
\begin{equation}
B(x)
=
\frac{c-\kappa(2x-1)}
     {a\xi\beta(1-ex)}.
\label{eq:behavioral_nullcline}
\end{equation}
Interior stationary states are intersections of $I=E(x)$ and $I=B(x)$. Boundary stationary states occur at $x=0$ or $x=1$, together with the corresponding epidemiological equilibrium.

\subsection*{Numerical analysis}

For the outcome maps in Fig.~\ref{fig:atlas}, we evaluated the model on a regular grid in the $(a,e)$ plane. At each parameter combination, we located all boundary and interior stationary states and classified them according to their local stability. We then recorded whether the stable set contained no adherence, full adherence, partial adherence, or combinations of these states.

Representative trajectories in Fig.~\ref{fig:dynamics} were obtained by integrating the full $(S,I,x)$ system from the indicated initial conditions. The trajectories shown in the $(x,I)$ plane are projections of the full three-dimensional dynamics. Parameter combinations at which an interior equilibrium lost stability were marked as oscillation. We show in the Supplementary Information that this loss of stability occurs through a Hopf bifurcation.

\subsection*{Acknowledgements}
This work was supported by the Japan Society for the Promotion of Science (JSPS) Grant-in-Aid for Scientific Research (KAKENHI) (JP26K24655 to R.K. and JP23KK0210, JP25K01465, JP25K22145 to H.I., JP22H04983, JP25K02340, JP26K02095 to M.Y.) and NIG-JOINT (94A2024, 51A2025, and 43A2026 to R.K. and 96A2023 and 73A2024 to H.I.).

\bibliographystyle{plainnat}
\bibliography{reference}

\end{document}


\maketitle

\setcounter{equation}{0}
\renewcommand{\theequation}{S\arabic{equation}}
\setcounter{figure}{0}
\renewcommand{\thefigure}{S\arabic{figure}}
\setcounter{table}{0}
\renewcommand{\thetable}{S\arabic{table}}

\section*{S1. Payoffs and classification of the instantaneous adherence game}

Individuals choose between adherence to a nonpharmaceutical intervention,
$A$, and non-adherence, $N$. Adherence entails a cost $c>0$.
At infection prevalence $I$, infection generates a perceived loss
$\xi\beta I$, where $\xi$ measures the perceived harm of infection and
$\beta$ is the baseline transmission rate.

Private protection reduces the susceptibility of an adherer by a fraction
$a$, whereas source control reduces transmission from an adherer to others
by a fraction $e$. Individuals additionally incur a social mismatch cost
$\kappa\geq 0$ when their behavior differs from that of their interaction
partner. The payoff matrix of the focal row player is therefore
\begin{equation}
\Pi(I) =
\bordermatrix{
      & N & A \cr
N     & -\xi\beta I
      & -\xi\beta I(1-e) -\kappa \cr
A     & -c-\xi\beta I(1-a)-\kappa
      & -c-\xi\beta I(1-a)(1-e) \cr
}.
\label{eq:S_payoff_matrix}
\end{equation}

When the population adherence frequency is $x$, random pairwise encounters
give the expected payoff to non-adherence,
\begin{align}
\pi_N(x,I)
&=
(1-x)(-\xi\beta I)
+x[-\xi\beta I(1-e)-\kappa]
\nonumber\\
&=
-\xi\beta I(1-ex)-\kappa x,
\label{eq:S_piN}
\end{align}
and the expected payoff to adherence,
\begin{align}
\pi_A(x,I)
&=
(1-x)[-c-\xi\beta I(1-a)-\kappa]
\nonumber\\
&\quad
+x[-c-\xi\beta I(1-a)(1-e)]
\nonumber\\
&=
-c-\xi\beta I(1-a)(1-ex)-\kappa(1-x).
\label{eq:S_piA}
\end{align}

The net incentive to adhere is defined by the difference between the payoffs:
\begin{align}
\alpha(x,I)
&=
\pi_A(x,I)-\pi_N(x,I)
\nonumber\\
&=
-c+a\xi\beta I(1-ex)+\kappa(2x-1).
\label{eq:S_alpha}
\end{align}
The first term is the cost of adherence. The second term is the private epidemiological benefit of adherence: private protection avoids a fraction $a$ of the infection loss that remains after source control supplied by other adherers. The third term is the frequency-dependent incentive generated by social matching.

Behavioral change follows
\begin{equation}
\frac{dx}{dt}
=
\epsilon x(1-x)\alpha(x,I),
\label{eq:S_replicator}
\end{equation}
where $\epsilon>0$ determines the rate of behavioral updating relative to epidemiological dynamics.

The signs of $\alpha(0,I)$ and $\alpha(1,I)$ classify the instantaneous game as non-adherence dominance, adherence dominance, chicken-like frequency dependence, or coordination-like frequency dependence, as summarized in Fig.~2A of the main text.

The endpoint incentives determine the prevalence thresholds for invasion and maintenance of adherence. When adherence is rare,
\begin{align}
    \alpha(0,I)=-c+a\xi\beta I-\kappa.
\end{align}
Rare adherers can increase when $\alpha(0,I)>0$. The corresponding invasion threshold is therefore
\begin{align}
    I_0=\frac{c+\kappa}{a\xi\beta},
\end{align}
for $a>0$.
When adherence is common,
\begin{align}
    \alpha(1,I)=-c+a(1-e)\xi\beta I+\kappa.
\end{align}
Full adherence can resist invasion by rare non-adherers when $\alpha(1,I)>0$. For $a>0$ and $e<1$, the corresponding maintenance threshold is
\begin{align}
    I_1=\frac{c-\kappa}{a(1-e)\xi\beta}.
\end{align}
When $\kappa>c$, $I_1<0$, meaning that full adherence can be maintained even in the absence of an epidemiological benefit.
The ordering of$I_0$ and $I_1$ determines the intermediate game structure. The two thresholds coincide when $I_0=I_1$,
which yields
\begin{align}
    e_c=\frac{2\kappa}{c+\kappa}.
    \label{eq:S_game_transition}
\end{align}
For $e<e_c$, $I_1<I_0$, producing an intermediate coordination-like region. For $e>e_c$, $I_0<I_1$, producing an intermediate chicken-like region.

At fixed $I$, the incentive is linear in $x$:
\begin{equation}
\alpha(x,I)
=
\alpha(0,I)
+x\left[2\kappa-ae\xi\beta I\right].
\label{eq:S_alpha_linear}
\end{equation}
When an interior mixed equilibrium exists, it is located at
\begin{equation}
x^\dagger
=
\frac{c+\kappa-a\xi\beta I}
     {2\kappa-ae\xi\beta I},
\label{eq:S_mixed_equilibrium}
\end{equation}
provided that the denominator is nonzero and $0<x^\dagger<1$.
The mixed equilibrium is attracting in the one-dimensional behavioral
dynamics when
\begin{equation}
2\kappa-ae\xi\beta I<0,
\label{eq:S_chicken_slope}
\end{equation}
as in the chicken-like game, and repelling when this quantity is positive,
as in the coordination game.


\section*{S2. Epidemic equilibria and socio-epidemiological stationary states}

For a fixed adherence frequency $x$, define
\begin{equation}
\rho=\frac{\beta}{\gamma+\mu},
\qquad
K=\frac{\delta+\mu}{\delta+\mu+\gamma}.
\label{eq:S_rhoK}
\end{equation}
The stable epidemiological equilibrium prevalence is
\begin{align}
E(x)
&=
\begin{cases}
K\left(1-\dfrac{1}{\rho b(x)}\right),
& \rho b(x) > 1,\\[6pt]
0,
& \rho b(x) \le 1,
\end{cases}
\\
b(x) &= (1-ax)(1-ex).
\end{align}


To connect the epidemiological and behavioral equilibria, we evaluate the adherence incentive along the epidemiological equilibrium curve $I=E(x)$. We therefore define
\begin{equation}
G(x)=\alpha(x,E(x)).
\label{eq:S_G}
\end{equation}
This function gives the net incentive to change adherence when infection prevalence is at the epidemiological equilibrium associated with a given adherence frequency $x$. An interior socio-epidemiological stationary state therefore satisfies
\begin{equation}
G(x^*)=0.
\label{eq:S_G_root}
\end{equation}
Solving this yields the interior behavioral nullcline
\begin{equation}
I
=
B(x)
=
\frac{c-\kappa(2x-1)}
     {a\xi\beta(1-ex)},
\label{eq:S_B}
\end{equation}
provided that $a>0$ and $1-ex>0$. Interior stationary states are
intersections of $I=E(x)$ and $I=B(x)$ in the biologically feasible region
$0<x<1$ and $I\geq0$. Cases with $a=0$ or $1-ex=0$ were evaluated directly
from Eq.~\eqref{eq:S_alpha}.

The derivative of the reduced feedback function is
\begin{equation}
G'(x)
=
a\xi\beta
\left[
(1-ex)E'(x)-eE(x)
\right]
+2\kappa.
\label{eq:S_Gprime}
\end{equation}
On the endemic branch,
\begin{equation}
E'(x)
=
\frac{K b'(x)}
     {\rho b(x)^2},
\qquad
b'(x)=-(a+e)+2aex.
\label{eq:S_Eprime}
\end{equation}
Because $b'(x)\leq0$ for $a,e,x\in[0,1]$, the endemic prevalence is
non-increasing with adherence.

Boundary stability follows directly from the invasion conditions: the no-adherence state is locally stable when
\begin{equation}
\alpha(0,E(0))<0,
\end{equation}
and the full adherence state is locally stable when
\begin{equation}
    \alpha(1,E(1))>0
\end{equation}


\section*{S3. Local stability of interior stationary states}

We next examine the local stability of an interior stationary state in the full socio-epidemiological system.
Because the system contains three dynamical variables, $S$, $I$, and $x$, stability cannot in general be inferred from the slope of the behavioral nullcline alone.
We therefore linearize the full system around an interior endemic stationary state $(S^*,I^*,x^*)$.

At such a stationary state,
\begin{equation}
\alpha(x^*,I^*)=0,
\qquad
\beta b(x^*)S^*=\gamma+\mu.
\label{eq:S_interior_conditions}
\end{equation}
Using these equilibrium conditions, the Jacobian is
\begin{equation}
J^*=
\begin{pmatrix}
-(\mu+\delta)-\beta b^*I^*
&
-(\delta+\gamma+\mu)
&
-\beta b^{\prime *}S^*I^*
\\[3pt]
\beta b^*I^*
&
0
&
\beta b^{\prime *}S^*I^*
\\[3pt]
0
&
\epsilon x^*(1-x^*)a\xi\beta(1-ex^*)
&
\epsilon x^*(1-x^*)
\left(2\kappa-ae\xi\beta I^*\right)
\end{pmatrix},
\label{eq:S_Jacobian_interior}
\end{equation}
where
\begin{equation}
b^*=b(x^*),
\qquad
b^{\prime *}=b'(x^*).
\end{equation}

To determine local stability, we consider the characteristic polynomial
\begin{equation}
\det(\lambda I-J^*)
=
\lambda^3+C_1\lambda^2+C_2\lambda+C_3.
\label{eq:S_characteristic}
\end{equation}
The coefficients are
\begin{align}
C_1
&=
-\left(J^*_{11}+J^*_{33}\right),
\label{eq:S_C1}\\
C_2
&=
-J^*_{12}J^*_{21}
+J^*_{11}J^*_{33}
-J^*_{23}J^*_{32},
\label{eq:S_C2}\\
C_3
&=
J^*_{11}J^*_{23}J^*_{32}
+J^*_{12}J^*_{21}J^*_{33}
-J^*_{13}J^*_{21}J^*_{32}.
\label{eq:S_C3}
\end{align}
The constant term $C_3$ has a useful interpretation.
Using the equilibrium conditions above together with Eq.~\eqref{eq:S_Gprime}, it can be rewritten as
\begin{equation}
C_3
=
-\epsilon x^*(1-x^*)
\beta b(x^*)I^*
(\mu+\delta+\gamma)
G'(x^*).
\label{eq:S_C3_Gprime}
\end{equation}
All factors multiplying $-G'(x^*)$ are positive at an interior endemic equilibrium.
Therefore, the condition $C_3>0$ requires
\begin{equation}
G'(x^*)<0.
\end{equation}
Thus, a negative slope of the reduced feedback function is necessary for local stability.
This condition alone is not sufficient, however, because stability depends on all three eigenvalues of the Jacobian.

The Routh--Hurwitz criterion gives the necessary and sufficient conditions for local asymptotic stability:
\begin{equation}
C_1>0,
\qquad
C_2>0,
\qquad
C_3>0,
\qquad
C_1C_2>C_3.
\label{eq:S_RH}
\end{equation}
An oscillatory loss of stability occurs when
\begin{equation}
C_1C_2=C_3,
\end{equation}
while $C_1,C_2,C_3>0$.
At this boundary, the characteristic polynomial factorizes as
\begin{equation}
(\lambda+C_1)(\lambda^2+C_2)=0,
\end{equation}
so that one eigenvalue remains negative whereas a complex-conjugate pair lies on the imaginary axis.
If this pair crosses the imaginary axis transversely as a parameter varies, the equilibrium undergoes a Hopf bifurcation.




\section*{S4. Conformity-driven instability and recurrent dynamics}

We examined the stability of the interior equilibrium for the parameter set corresponding to the recurrent dynamics in Fig.~3B. At $\epsilon=1$, the equilibrium is stable under weak conformity but loses stability at $\kappa\simeq0.55$ (Fig.~S1A,B). Beyond this threshold, a stable periodic orbit emerges. The oscillation amplitude increases with conformity, eventually producing large fluctuations between low and high adherence and corresponding variation in infection prevalence.

For each value of $\kappa$, we located the interior stationary state by solving $G(x)=0$ and evaluated its local stability from the eigenvalues of the full Jacobian. Periodic-orbit extrema were obtained by numerically integrating the full $(S,I,x)$ system after discarding an initial transient and recording the minimum and maximum values over the subsequent time interval. The stability boundary in Fig.~S1C was located from the condition that the largest real part of the Jacobian eigenvalues changes sign.

The conformity threshold for instability depends on the behavioral updating rate (Fig.~S1C). When behavioral adjustment is slow, the interior equilibrium remains stable over a wider range of conformity. Faster behavioral adjustment lowers the conformity required to destabilize the equilibrium. As $\epsilon$ increases, the stability boundary approaches
\begin{equation}
\kappa=\frac{ce}{2-e},
\end{equation}
the threshold at which the intermediate instantaneous game changes between chicken-like and coordination-like structures (dashed line). For this parameter regime, conformity provides the positive frequency dependence required for instability, whereas the behavioral updating rate determines how readily this social reinforcement destabilizes the coupled epidemiological equilibrium.

\begin{figure}
    \centering
    \includegraphics[width=1\linewidth]{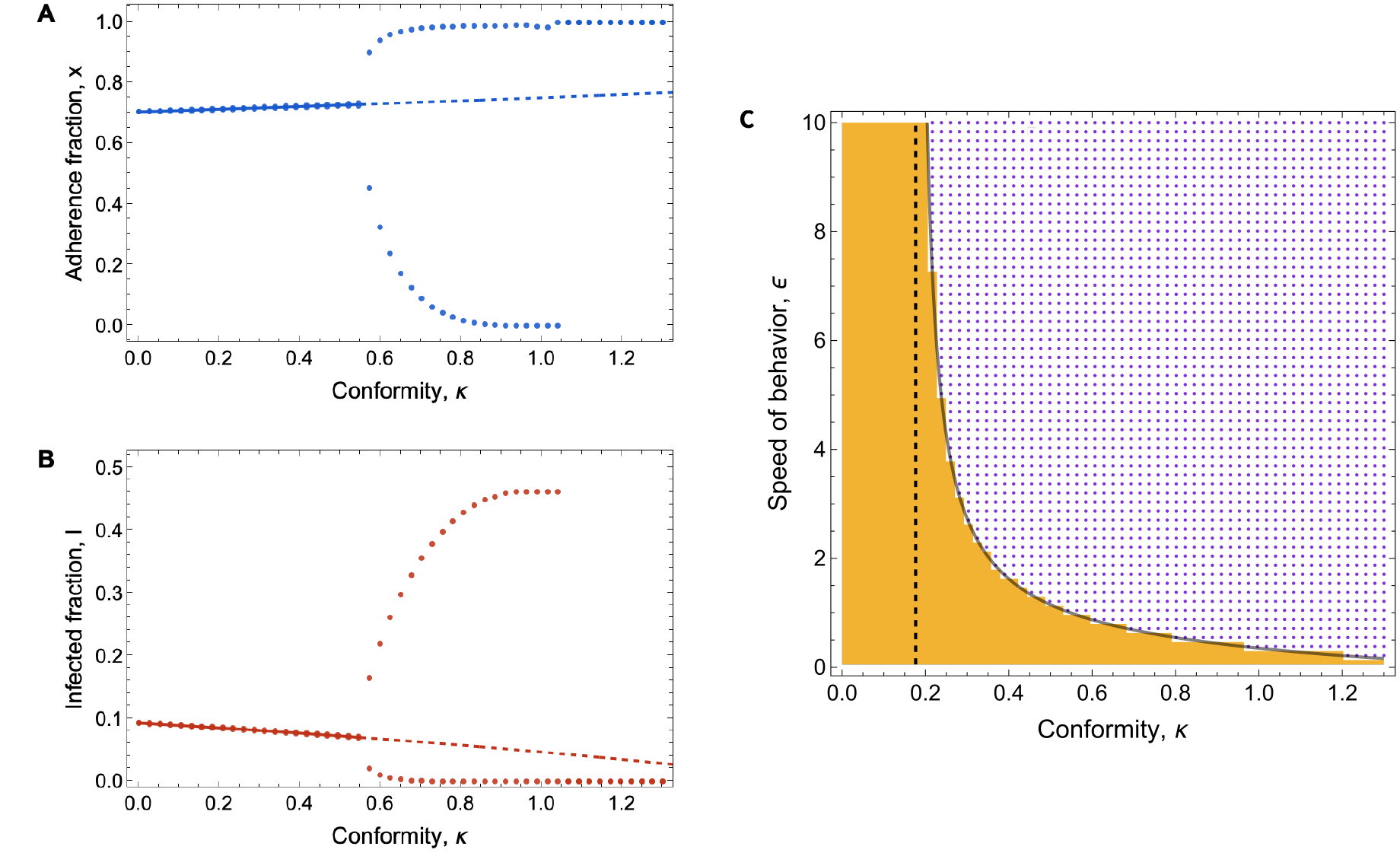}
    \caption{\textbf{Conformity and behavioral updating determine the stability of the interior socio-epidemiological equilibrium.} (A,B) Interior-equilibrium values and periodic-orbit extrema as conformity strength $\kappa$ varies at $\epsilon=1$. Solid and dashed curves denote locally stable and unstable interior equilibria, respectively. Points show the minimum and maximum values of (A) adherence $x$ and (B) infection prevalence $I$ along the stable periodic orbit. (C) Local stability of the interior equilibrium in the $(\kappa,\epsilon)$ plane. Yellow shading denotes a stable equilibrium, whereas purple dotted hatching denotes an unstable equilibrium. The solid curve marks the oscillatory-instability boundary. The vertical dashed line indicates $\kappa^*=ce/(2-e)$, which is approached by the stability boundary as $\epsilon\to\infty$. Fixed parameters are $\beta=4$, $\gamma=0.98$, $\mu=0.02$, $\delta=0.782$, $c=1$, $\xi=4$, $a=0.85$, and $e=0.30$.
}
\end{figure}

\begin{figure}
    \centering
    \includegraphics[width=0.7\linewidth]{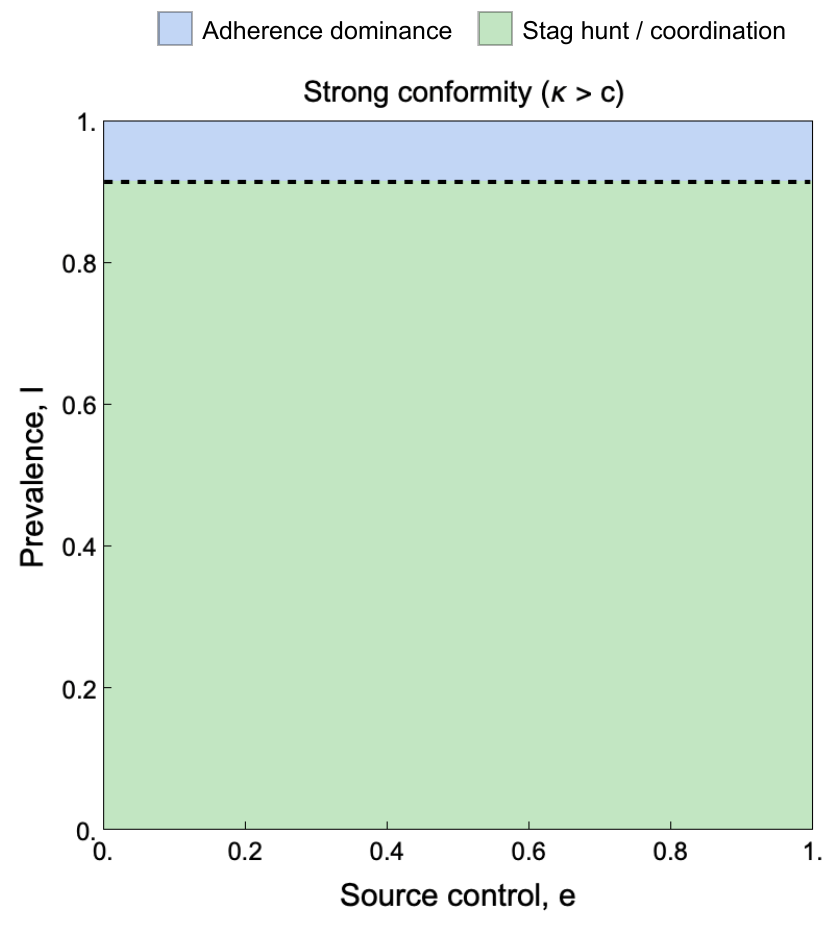}
    \caption{\textbf{Instantaneous game structure under strong conformity.} Instantaneous game structure across source control $e$ and prevalence $I$ when conformity exceeds the cost of adherence ($\kappa>c$). Colors indicate the game types defined in Fig. 2A. Because $\kappa>c$, adherence can always be maintained when common, and the game is always either coordination-like or adherence-dominant. The horizontal dashed boundary satisfies $\alpha(0,I)=0$. Parameters are $a\xi\beta/c=2.4$ and $\kappa/c=1.2$.
}
\end{figure}

\begin{figure}
    \centering
    \includegraphics[width=1\linewidth]{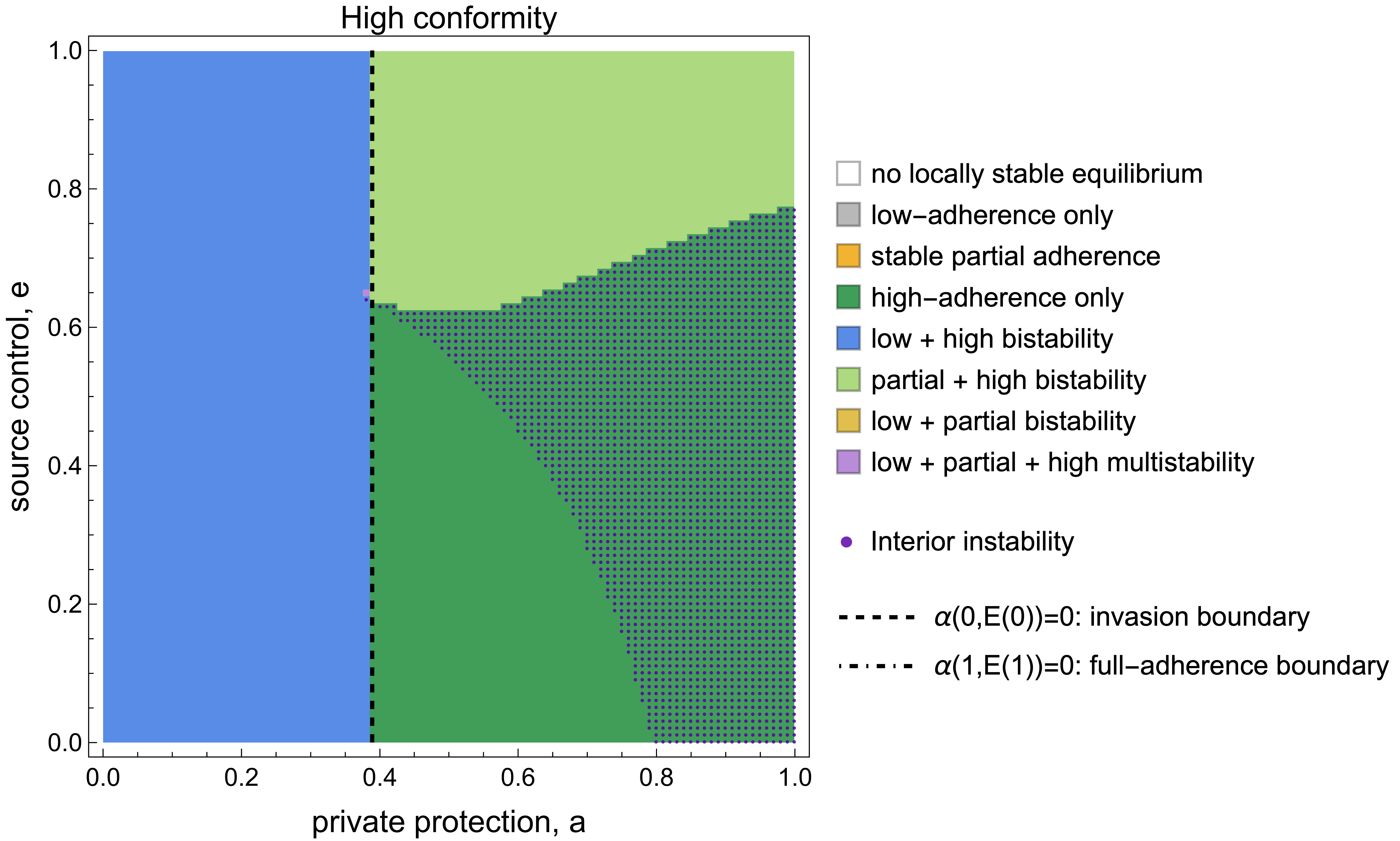}
    \caption{\textbf{Strong conformity reorganizes the landscape of stable socio-epidemiological states.} Locally stable stationary states across the private-protection–source-control plane under strong conformity, $\kappa=1.2>c$. Colors and patterns indicate the stable state or combination of stable states, as defined in the legend; gray denotes parameter combinations with no locally stable stationary state detected. The dashed black line marks the change in stability of the no-adherence state, defined by $\alpha(0,E(0))=0$. Because $\kappa>c$, the full-adherence state is locally stable throughout the parameter plane and no $\alpha(1,E(1))=0$ boundary occurs. Purple points indicate parameter combinations at which the interior equilibrium is unstable. Fixed parameters are $\beta=4.0$, $\gamma=0.98$, $\mu=0.02$, $\delta=0.782$, $c=1.0$, $\xi=4$, and $\epsilon=1.0$.
}
\end{figure}